\documentclass[twocolumn,showpacs,amsmath,amssymb,aps,prl]{revtex4}

\usepackage{bm}

\newcommand{\itGamma}{{\mathit{\Gamma}}}
\newcommand{\del}{\partial}

\newcommand{\cT}{{\cal T}}
\newcommand{\cR}{{\cal R}}

\newcommand{\cfl}[2]{{\textstyle {{#1}\brace {#2}}}}
\newcommand{\ds}{\displaystyle}
\newcommand{\lc}{\varepsilon}
\newcommand{\vect}[1]{\mathbf{#1}}

\begin{document}

\title{Dirac particle in gravitational field}

\author{Milovan Vasili\'c}
\email{mvasilic@phy.bg.ac.yu}
\author{Marko Vojinovi\'c}
\email{vmarko@phy.bg.ac.yu}
\affiliation{Institute of Physics, P.O.Box 57, 11001 Belgrade, Serbia}


\begin{abstract}
Classical dynamics of spinning zero-size objects in an external
gravitational field is derived from the conservation law of the
stress-energy and spin tensors. The resulting world line equations differ
from those in the existing literature. In particular, the spin of the Dirac
particle does not couple to the background curvature. As a check of
consistency, the wave packet solution of the free Dirac equation is
considered. The resulting equations are shown to include a constraint that
relates the wave packet spin to its orbital angular momentum. In the
zero-size limit, both contributions to the total angular momentum disappear
simultaneously.
\end{abstract}


\maketitle

\textbf{Introduction.} The problem of particle motion in external
gravitational field is usually addressed by using the Mathisson-Papapetrou
method \cite{Mathisson1937, Papapetrou1951}. One starts with the covariant
conservation law of the stress-energy tensor of matter fields, and analyzes it
under the assumption that matter is highly localized in space. In the lowest,
single-pole approximation, the moving matter is viewed as a point particle. In
the pole-dipole approximation, its non-zero size is taken into account.

The results found in literature can be summarized as follows. Spinless
particles in the single-pole approximation obey the geodesic equation. In
the pole-dipole approximation, the rotational angular momentum of the
localized matter couples to spacetime curvature, and produces geodesic
deviations \cite{Mathisson1937, Papapetrou1951, Tulczyjew1959, Taub1964,
Dixon1964, Dixon1970}. If the particles have spin, the curvature couples to
the total angular momentum \cite{Trautman1972, Hehl1976a, Yasskin1980,
Nomura1991, Nomura1992}.

What we are interested in is a consistent single-pole analysis of spinning
particles in an external gravitational field. This is motivated by the
observation that single-pole approximation eliminates the influence of
particle thickness, and allows the derivation of the pure spin-curvature
coupling. In fact, this is the only way to see the influence of curvature on
the spin part of the total angular momentum. The ambiguous algebraic
decomposition of the total angular momentum into spin and orbital
contributions are of no help. What we need is a truly zero-size object. The
existing literature on the subject turns out not to have this sort of
prediction.

The results that we have obtained are summarized as follows. Trajectories of
spinning zero-size massive particles generally deviate from the geodesic
lines. The deviation is due to the spin-curvature and spin-torsion
couplings. These turn out to be different from what has been believed so
far. In particular, the spin of the free Dirac point particle does not
couple to the curvature. If it is viewed as a wave packet solution of the
Dirac equation, it does not couple to the torsion either. The Dirac point
particles behave as spinless objects.

In what follows, we shall outline our derivation, and make the necessary
comments. We begin with the covariant conservation of the fundamental matter
currents---stress-energy tensor $\tau^{\mu}{}_{\nu}$ and spin tensor
$\sigma^{\lambda}{}_{\mu\nu}$:
\begin{subequations} \label{jna1}
\begin{eqnarray}
&\ds\left( D_{\rho} + \cT^{\lambda}{}_{\rho\lambda} \right)
\tau^{\rho}{}_{\mu} =
\tau^{\nu}{}_{\rho}\cT^{\rho}{}_{\mu\nu} + \frac{1}{2}
\sigma^{\nu}{}_{\lambda\rho}\cR^{\lambda\rho}{}_{\mu\nu}\,,&   \label{jna1a} \\
&\ds\left( D_{\rho}+\cT^{\lambda}{}_{\rho\lambda}\right)
\sigma^{\rho}{}_{\mu\nu}=\tau_{\mu\nu} - \tau_{\nu\mu}\,.&     \label{jna1b}
\end{eqnarray}
\end{subequations}
Here, $\cT^{\lambda}{}_{\mu\nu}$ and $\cR^{\lambda\rho}{}_{\mu\nu}$ are the
spacetime torsion and curvature, and $D_{\mu}v^{\nu} \equiv \del_{\mu} v^{\nu}
+ \itGamma^{\nu}{}_{\rho\mu} v^{\rho}$ is the covariant derivative with the
nonsymmetric connection $\itGamma^{\rho}{}_{\mu\nu}$. The second equation
clearly shows that the antisymmetric part of the stress-energy tensor is not a
dynamic variable. Indeed, given the spin tensor $\sigma^{\lambda\mu\nu}$, the
antisymmetric part $\tau^{[\mu\nu]}$ is completely determined. We can use the
equation (\ref{jna1b}) to eliminate $\tau^{[\mu\nu]}$ components, and rewrite
the conservation equations in terms of $\tau^{(\mu\nu)}$ and
$\sigma^{\lambda\mu\nu}$ alone.

Let us briefly explain how the assumption of highly localized matter
simplifies the conservation equations (\ref{jna1}). First, we expand the
stress-energy and spin tensors of the localized matter in multipoles around
a suitably chosen timelike line $x^{\mu}=z^{\mu}(\tau)$. Using the results of
Ref. \cite{Vasilic2007}, a spacetime function $F(x)$ is written as a series of
$\delta$-function derivatives:
\begin{equation} \label{jna2}
\begin{array}{cl}
\ds F(x) = \int d\tau \bigg[ & \ds M(\tau)
\frac{\delta^{(4)}(x-z)}{\sqrt{-g}}                          \\
& \ds - \nabla_{\rho}\left( M^{\rho}(\tau)
\frac{\delta^{(4)}(x-z)}{\sqrt{-g}} \right) + \cdots \bigg]. \\
\end{array}
\end{equation}
The line parameter $\tau$ is chosen to be the proper distance, $d\tau^2 =
g_{\mu\nu}dz^{\mu}dz^{\nu}$, and $\nabla_{\rho}\equiv D_{\rho}(\itGamma\to
\{\})$ is the Riemannian covariant derivative. Here, $\cfl{\lambda}{\mu\nu}$
stands for the Levi-Civita connection. The expansion (\ref{jna2}) is
manifestly covariant, with the multipole coefficients $M^{\rho_1\dots
\rho_n}$ being tensors with respect to spacetime diffeomorphisms. The
adopted metric signature is $+2$, and $g\equiv \det (g_{\mu\nu})$ is the
metric determinant.

The decomposition (\ref{jna2}) is suitable for treating matter which is well
localized around the line $x^{\mu}=z^{\mu}(\tau)$. In fact, the variable
$F(x)$ must drop exponentially to zero as we move away from the line if we
want the series (\ref{jna2}) to be well defined. If this is the case, the
coefficients $M^{\rho_1\dots\rho_n}$ get smaller as $n$ gets larger. In the
lowest, single-pole approximation, all higher $M$'s are neglected. In the
pole-dipole approximation one neglects all but the first two
$M$-coefficients.

\textbf{World line equations.} We expand the stress-energy and spin tensors
of the localized matter in multipoles around the line
$x^{\mu}=z^{\mu}(\tau)$. Being concerned with the case of point particles,
we choose to work in the single-pole approximation. Thus,
\begin{subequations} \label{jna3}
\begin{eqnarray}
\tau^{(\mu\nu)} & = & \ds \int d\tau \, c^{\mu\nu}(\tau)
\frac{\delta^{(4)}(x-z)}{\sqrt{-g}} \,, \label{jna3a} \\
\sigma^{\lambda\mu\nu} & = & \ds \int d\tau \, c^{\lambda\mu\nu}(\tau)
\frac{\delta^{(4)}(x-z)}{\sqrt{-g}} \,. \label{jna3b}
\end{eqnarray}
\end{subequations}
This definition of single-pole approximation differs from what is found in
the existing literature \cite{Hehl1976a, Nomura1991}. There, the
antisymmetric part of the stress-energy tensor $\tau^{[\mu\nu]}$ had an
equal treatment as $\tau^{(\mu\nu)}$. As $\tau^{[\mu\nu]}$ is not an
independent variable, this imposed unnecessary constraints on
$\sigma^{\lambda\mu\nu}$. In particular, the spin of the Dirac particle was
ruled out. To overcome this problem, the authors of Ref. \cite{Nomura1991}
abandoned single-pole in favour of pole-dipole approximation. Their
subsequent limit of vanishing orbital angular momentum brought them a
non-trivial spin-curvature coupling, even in the Dirac case. In what
follows, we shall demonstrate that it is not quite so.

Let us substitute (\ref{jna3}) in (\ref{jna1}), and solve the obtained
equations in terms of the unknown variables $z^{\mu}(\tau)$,
$c^{\mu\nu}(\tau)$ and $c^{\lambda\mu\nu}(\tau)$. The procedure is very much
the same as described in Ref. \cite {Vasilic2007}. First, we eliminate the
equation (\ref{jna1b}), and rewrite (\ref{jna1a}) in terms of the of the
independent variables $\tau^{(\mu\nu)}$ and $\sigma^{\lambda\mu\nu}$. We
then multiply (\ref{jna1a}) with an arbitrary vector valued function
$f_{\mu}(x)$, and integrate over the spacetime. The resulting equation
depends on the function $f_{\mu}$ and its first and second covariant
derivatives, evaluated on the line $x^{\mu}=z^{\mu}(\tau)$. Owing to the
arbitrariness of the function $f_{\mu}(x)$, the terms proportional to its
independent derivatives separately vanish. The obtained equations are
manifestly covariant with respect to spacetime diffeomorphisms.

Skipping the details of the calculation, which will be published elsewhere,
we present here the final result. It consists of two sets of world line
equations. The first is a set of algebraic equations which determine the
coefficients $c^{\mu\nu}$ and $c^{\lambda\mu\nu}$ in terms of the free
parameters $m(\tau)$, $s^{\mu\nu}(\tau)$ and $s^{\lambda\mu\nu}(\tau)$:
\begin{subequations} \label{jna4}
\begin{eqnarray}
&\ds c^{\mu\nu}=m u^{\mu}u^{\nu}+2u_{\lambda}u^{(\mu}\nabla s^{\nu)\lambda}-
     \frac{1}{2}K_{\lambda\rho}{}^{(\mu} c^{\nu)\lambda\rho},&    \label{jna4a} \\
&\ds c^{\lambda\mu\nu}=2u^{\lambda}s^{\mu\nu}+s^{\lambda\mu\nu}.& \label{jna4b}
\end{eqnarray}
\end{subequations}
Here, $u^{\mu}\equiv dz^{\mu}/d\tau$ is the particle $4$-velocity,
$K_{\mu\lambda\nu} \equiv
(\cT_{\nu\mu\lambda}-\cT_{\mu\lambda\nu}-\cT_{\lambda\nu\mu})/2$ is the
contorsion tensor, and $\nabla$ stands for the Riemannian covariant
derivative along the particle trajectory ($\nabla v^{\mu}=dv^{\mu}/d\tau
+\cfl{\mu}{\lambda\rho}v^{\lambda}u^{\rho}$). The parameters $s^{\mu\nu}$
and $s^{\lambda\mu\nu}$ are totally antisymmetric tensors:
$$
s^{\mu\nu} = - s^{\nu\mu}, \qquad
s^{\lambda\mu\nu} = - s^{\lambda\nu\mu} = s^{\nu\lambda\mu}.
$$
The algebraic equations (\ref{jna4}) differ from what we find in literature
by the form of (\ref{jna4b}).

The second set of equations are the differential equations that determine
the particle world line and spin precession. They read:
\begin{subequations} \label{jna5}
\begin{eqnarray}
\nabla p^{\mu}-u^{\nu}s^{\rho\lambda} R^{\mu}{}_{\nu\rho\lambda} & = & \ds
\frac{1}{2}c_{\nu\rho\lambda}\nabla^{\mu} K^{\rho\lambda\nu},\label{jna5a} \\
\nabla s^{\mu\nu} + D^{\mu\nu} & = &  u^{[\mu}p^{\nu ]} , \label{jna5b}
\end{eqnarray}
\end{subequations}
where
\begin{eqnarray}
p^{\mu} & \equiv &  mu^{\mu} + 2 u_{\rho}\left( \nabla s^{\mu\rho} +
D^{\mu\rho} \right) , \nonumber \\
D^{\mu\nu} & \equiv &  c^{\rho\lambda[\nu} K^{\mu]}{}_{\lambda\rho} +
\frac{1}{2} K_{\lambda\rho}{}^{[\mu} c^{\nu]\rho\lambda} . \nonumber
\end{eqnarray}
Here, the symbol $R^{\mu}{}_{\nu\rho\lambda}$ stands for the Riemann
curvature, $R^{\mu}{}_{\nu\rho\lambda}\equiv
\cR^{\mu}{}_{\nu\rho\lambda}(\itGamma\to \{\})$, and both external fields,
$R^{\mu}{}_{\nu\rho\lambda}$ and $K^{\rho\lambda\nu}$, are evaluated on the
line $x^{\mu}=z^{\mu}(\tau)$.

The obtained single-pole equations differ from the known pole-dipole result
\cite{Trautman1972, Hehl1976a, Yasskin1980, Nomura1991, Nomura1992} by the
presence of the constraint (\ref{jna4b}). It is a consequence of our
assumption that the particle has no thickness, and therefore no orbital
degrees of freedom. In the existing literature, an analogous but more
restrictive constraint appears in this regime \cite{Hehl1976a, Nomura1991}.
This is because the antisymmetric part of the stress-energy tensor
$\tau^{[\mu\nu]}$ has been treated as an independent variable, in spite of
the restriction (\ref{jna1b}). In our approach, the only independent
variables are $\sigma^{\lambda\mu\nu}$ and $\tau^{(\mu\nu)}$, and the
resulting constraint (\ref{jna4b}) is not so strong. In particular, it does
not rule out the free Dirac field, or any other massive elementary field.
Indeed, the formula $c^{\lambda\mu\nu} = u^{\lambda}s^{\mu\nu} +
\frac{1}{s}u^{[\mu}s^{\nu]\lambda}$ for the spin tensor of the elementary
particle of spin $s$ (see Ref. \cite{Nomura1992}) is a special case of
(\ref{jna4b}).

In what follows, we shall examine the special case of spin $1/2$ pointlike
matter. Surprisingly, we shall discover that spin $1/2$ does not couple to
the curvature, leading to geodesic trajectories in torsionless spacetimes.

\textbf{Dirac particle.} The basic property of Dirac matter is the total
antisymmetry of its spin tensor $\sigma^{\lambda\mu\nu}$. As a consequence,
the coefficients $c^{\lambda\mu\nu}$ are also totally antisymmetric, and the
constraint (\ref{jna4b}) implies
$$
s^{\mu\nu}=0\,.
$$
The vanishing of the $s^{\mu\nu}$ component of the spin tensor has far
reaching consequences. First, we see that the spin-curvature coupling
disappears from the world line equation (\ref{jna5a}). Second, the spin
precession equation (\ref{jna5b}) becomes an algebraic equation. If we
define the spin vector $s^{\mu}$ by $s^{\mu\nu\rho}\equiv
e^{\mu\nu\rho\lambda}s_{\lambda}$, and the axial component of the contorsion
$K^{\mu}$ as $K^{\mu} \equiv e^{\mu\nu\rho\lambda} K_{\nu\rho\lambda}$, the
equations (\ref{jna5}) become
\begin{subequations} \label{jna6}
\begin{eqnarray}
& \ds\nabla\left( mu^{\mu}+K^{[\mu}s^{\nu]}u_{\nu} \right)+
  \frac{1}{2} s^{\nu}\nabla^{\mu}K_{\nu} =0 \,, &            \label{jna6a} \\
& \ds K_{\perp}^{[\mu}s_{\perp}^{\nu]} = 0\,. &              \label{jna6b}
\end{eqnarray}
\end{subequations}
Here, $e_{\mu\nu\rho\lambda} \equiv \sqrt{-g}\,\lc_{\mu\nu\rho\lambda}$ is
the covariant Levi-Civita symbol, and the decomposition of a vector into
components parallel and orthogonal to the world line is used: $v^{\mu}
\equiv v_{\perp}^{\mu} + vu^{\mu}$, $v_{\perp}^{\mu}u_{\mu} \equiv 0$. As we
can see, the spin couples only to the axial component of the contorsion,
which means that \emph{Dirac point particles follow geodesic trajectories in
torsionless spacetimes}. At the same time, the absence of torsion
trivializes the equation (\ref{jna6b}), and no information on the behavior
of the spin vector is available.

To make it clear, only zero-size Dirac particles lack the spin-curvature
coupling. In the case of thick particles (pole-dipole approximation),
curvature couples to the total angular momentum, thus to both, spin and
orbital angular momentum. Therefore, our result does not compromise the
known pole-dipole result. It just explains its behaviour in the zero-size
limit.

If the background torsion has nontrivial axial component, a geodesic
deviation appears, but also a very strong constraint on the spin vector.
Indeed, the equation (\ref{jna6b}) implies that the orthogonal component of
$s^{\mu}$ always orients itself along the background direction
$K_{\perp}^{\mu}$. This unusual behavior suggests that the spin vector of
the Dirac point particle might be zero, after all. In fact, the world line
equations (\ref{jna4}) and (\ref{jna5}) are derived under very general
assumptions of the existence of pointlike solutions in an arbitrary field
theory. They do not care about peculiarities of specific theories, or
specific types of localized solutions. What might happen in a particular
theory is that additional constraints appear to ensure the existence of
localized solutions. In what follows, we shall analyze the wave packet
solutions of the flat space Dirac equation, with the idea to check if they
can be viewed as point particles. We shall see that this is possible only if
the particle spin goes to zero.

\textbf{Wave packet.} Wave packet is a solution which is well localized in
space, but resembles a plane wave inside. To be viewed as a particle, its
size $\ell$ is considered in the limit $\ell\to 0$. At the same time, the
particle stability is achieved in the limit $\lambda / \ell \to 0$, where
$\lambda$ is its wavelength. We shall consider the free Dirac equation, and
construct wave packets as follows. In the initial moment $t=0$, we choose
the configuration $\psi(\vect{r},0) \equiv g(\vect{r})\psi_p(\vect{r},0)$,
where $\psi_p(x) \equiv a(k)\exp (ik_{\mu}x^{\mu})$ is a plane wave solution
of the Dirac equation, and $g(\vect{r}) \sim \exp (-r^2/\ell^2)$ is a
function that cuts out a small piece of the plane wave. Using the Dirac
equation, we can calculate time derivatives, and thereby determine its time
evolution. In fact, we only need first time derivatives, as neither
$\tau^{\mu\nu}$ nor $\sigma^{\lambda\mu\nu}$ depend on higher derivatives:
\begin{equation} \label{jna7}
\tau_{\mu\nu} = i \bar{\psi} \gamma_{\mu}
\mathord{\stackrel{\leftrightarrow}{\del}}{}_{\nu}\psi \,, \qquad
\sigma^{\lambda\mu\nu} = \lc^{\lambda\mu\nu\rho} \bar{\psi} \gamma_5
\gamma_{\rho} \psi \,.
\end{equation}
In addition, there is a conserved current $j^{\mu} =
\bar{\psi}\gamma^{\mu}\psi$, associated with the extra $U(1)$ symmetry of the
Dirac action. The wave packet expressions of these currents are obtained
straightforwardly, but we shall skip these details. Instead,
we shall discuss two interesting relations that we have
discovered to further constrain the currents. The first,
\begin{equation} \label{jna8}
\lc_{\mu\nu\rho\lambda} \sigma^{\mu\nu\rho} j^{\lambda} =0 \,,
\end{equation}
shows that the spin vector $s^{\mu}$ is orthogonal to $u^{\mu}$, as expected.
This follows from the fact that $j^{\mu} \propto u^{\mu}$ in the single-pole
approximation. The second relation,
\begin{equation} \label{jna9}
\tau^{\mu\nu}j_{\nu} \propto j^{\mu},
\end{equation}
has even more striking consequences. Its physical meaning is best seen in
the rest frame where it takes the form $\tau^{\alpha 0} = j^{\alpha}=0$. It
tells us that the two currents are mutually proportional, i.e. that the
energy and charge flow in the same direction. In the limit $\lambda / \ell
\to 0$, one can show that it implies the constraint
\begin{equation} \label{jna10}
x^{\alpha}\tau^{(0\beta)} - x^{\beta}\tau^{(0\alpha)}
= \sigma^{0\alpha\beta} + div \,,
\end{equation}
where $div$ stands for the total $3$-divergence. In the single-pole
approximation, only the space integral of this relation is needed. Then, the
divergence term disappears, and we end up with the constraint that relates
the wave packet spin to its orbital angular momentum. As orbital degrees of
freedom are neglected in this approximation, so is the spin. Thus, the
particle {\it orbital angular momentum and spin disappear simultaneously} in
the limit $\ell\to 0$, $\lambda /\ell \to 0$.

To summarize, we see that the spin vector $s^{\mu}$ vanishes in the single-pole
approximation, and the particle trajectory becomes a geodesic line even in the
presence of torsion. This is a valid conclusion only if Dirac point particles are
viewed as zero-size wave packets. It does not apply to other localized field
configurations. For example, one could consider a localized bound state of Dirac
and electromagnetic fields. Its spin-tensor is still totally antisymmetric (which
implies the vanishing spin-curvature coupling), but nothing proves its
disappearance in the zero-size limit. The validity of the above conclusion also
requires some sort of equivalence principle to hold. This is because the
considered wave packet is a solution of the free Dirac equation, and the
inclusion of curvature or torsion may destroy it. What we can do is to consider
weak gravity, so that terms quadratic in curvature and torsion are neglected. In
that case, the free wave packets are a good approximation to the exact solution,
which implies that {\it Dirac point particles behave as spinless objects in a
weak gravitational field}. They can still probe the spacetime curvature, but for
the probe of the background torsion, one needs a thick particle.

Let us note here that one could use the alternative WKB approach to study
Dirac field in curved backgrounds. In such an approach, the starting point
is the covariant Dirac equation rather than the general conservation laws.
However, whatever solution of the covariant Dirac equation is used, it must
obey the covariant conservation equations (\ref{jna1}). This is because the
conservation laws (\ref{jna1}) are a general property of all covariant
Lagrangians, the covariant Dirac Lagrangian being just one of those. Once
we check that the same approximation is used in both approaches, the two
predictions must not contradict each other. In particular, if the wave
packet solution of the covariant Dirac equation is considered in the
zero-size limit, its trajectory will belong to the class of trajectories
defined by our world line equations. The WKB analysis of the covariant Dirac
equation has already been considered in literature (see, for example, Ref.
\cite{Audretsch}). However, this kind of analysis is not suitable for the
zero-size limit, as the considered wave configuration is far from resembling
a point-particle. What is called particle trajectory in this approach is
rather a ray of geometric optics.

Let us close our exposition with the speculation that Dirac particle is not
the only one that possesses these features. Indeed, the relations analogous
to (\ref{jna9}) might exist quite generally. There is nothing special about
the statement that all the particle charges flow in the same direction. This
is something one would expect to hold for any type of localized matter.
However, the consequence (\ref{jna10}) has been derived for the Dirac wave
packet solution only. If this turns out to be the general property of wave
packet solutions, the disappearance of spin in the zero-size limit might be
a feature of all massive point particles.

\begin{acknowledgments}
This work was supported by the Serbian Science Foundation, Serbia.
\end{acknowledgments}

\end{document}